\documentclass[preprint2]{aastex}

\begin{document}

\title{Stochastic Processes in Yellow and Red Pulsating Variables}

\author{David G. Turner}
\affil{Saint Mary's University,
    Halifax, Nova Scotia, Canada}
\email{turner@ap.smu.ca}

\author{J. R. Percy, T. Colivas}
\affil{University of Toronto, Toronto, Ontario, Canada}

\author{L. N. Berdnikov}
\affil{Sternberg Astronomical Institute, Moscow, Russian Federation}
 
\author{M. Abdel-Sabour Abdel-Latif}
\affil{National Research Institute of Astronomy and Geophysics, Helwan, Cairo, Egypt}

\begin{abstract}
Random changes in pulsation period are well established in cool pulsating stars, in particular the red giant variables: Miras, semi-regulars of types A and B, and RV Tau variables. Such effects are also observed in a handful of Cepheids, the SX Phe variable XX Cyg, and, most recently, the red supergiant variable, BC Cyg, a type C semi-regular. The nature of such fluctuations is seemingly random over a few pulsation cycles of the stars, yet the regularity of the primary pulsation mechanism dominates over the long term. The degree of stochasticity is linked to the dimensions of the stars, the randomness parameter {\it e} appearing to correlate closely with mean stellar radius through the period {\it P}, with an average value of $e/P = 0.0136 \pm0.0005$. The physical processes responsible for such fluctuations are uncertain, but presumably originate in temporal modifications of envelope convection in such stars.
\end{abstract}

\section{Introduction}
It is well known in variable star observations that it is almost impossible to predict the exact moments of light maximum in some late-type pulsating variables, such as Miras and semi-regular variables. The cyclical light patterns displayed in such stars are reasonably well defined over long time intervals, but the regularity of their pulsation is typically marked by other effects best revealed through careful O--C analysis. A common effect is that of ``random'' fluctuations in pulsation period for a star from one cycle to another. Eighty years ago \citet{1} developed a novel method for establishing the importance of such random fluctuations for Mira variables, and the technique has been revived frequently in recent years \citep{2,3,4,5,6} in order to establish its importance in other Mira variables as well as in other types of pulsating variables.

\begin{figure}[h]
\includegraphics[width=0.4\textwidth]{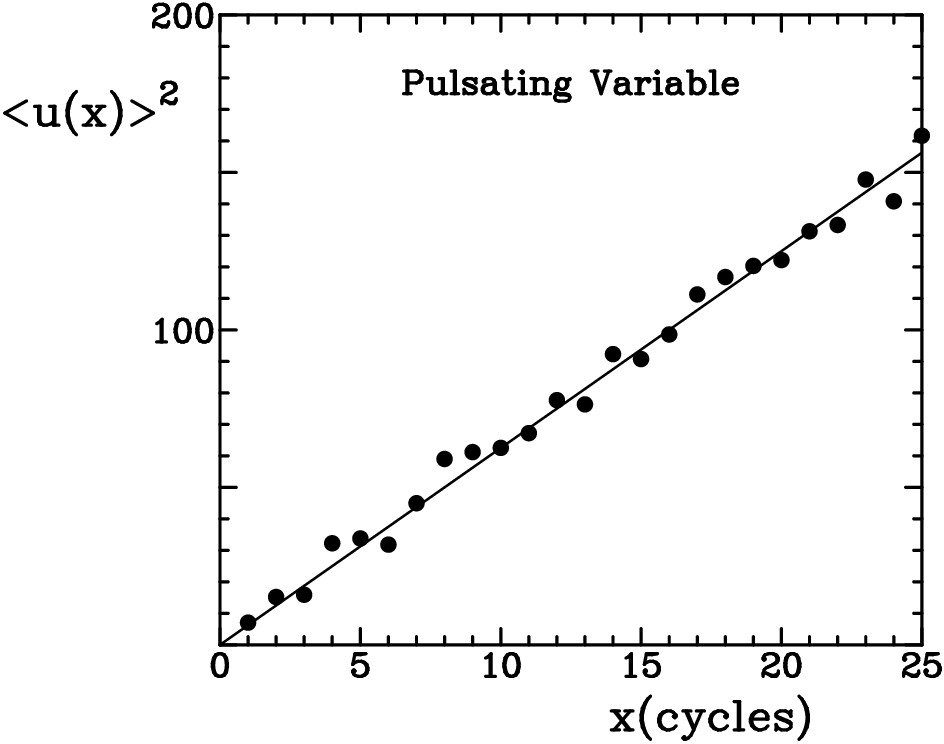}
\caption{\label{fig1} Schematic expectations for random period variations in a pulsating star with randomness factor $e = 2.5$ and uncertainties of $a = 0^{\rm d}.08$ in determining times of light maximum, according to predictions by \citet{1}.}
\end{figure}

In essence the technique involves computing, without regard to sign, the average accumulated time delays $\langle u(x) \rangle$ between light maxima separated by {\it x} cycles. If the time deviations in light maxima are dominated by random fluctuations in pulsation period, then as demonstrated by \citet{1} the data for all available observed light maxima should display a trend described by:
\begin{displaymath}
\langle u(x) \rangle^2 = 2 a^2 + x e^2 \;,
\end{displaymath}
where {\it a} represents the average uncertainty in days for established times of light maxima and {\it e} represents the magnitude of any random fluctuations in period. Fig.~\ref{fig1} represents schematically such expectations for a pulsating variable with a randomness factor of $e = 2.5$ and uncertainties of $a = 0.08$ day in established times of light maxima.

\section{Results for Different Pulsators}
Convincing evidence for random cycle-to-cycle variations in pulsation period for several different types of long-period variables has been gathered by Percy and collaborators \citep{2,3,4} based upon the Eddington-Plakidis method. In contrast, mostly negative findings were obtained for shorter-period pulsators such as the $\beta$ Cep star BW Vul \citep{5} and a sample of $\delta$ Sct and RR Lyr variables \citep{6}, an exception being the SX Phe variable XX Cyg, for which a weak signal of random fluctuations in period was obtained \citep{6}.

\begin{figure}[h]
\includegraphics[width=0.45\textwidth]{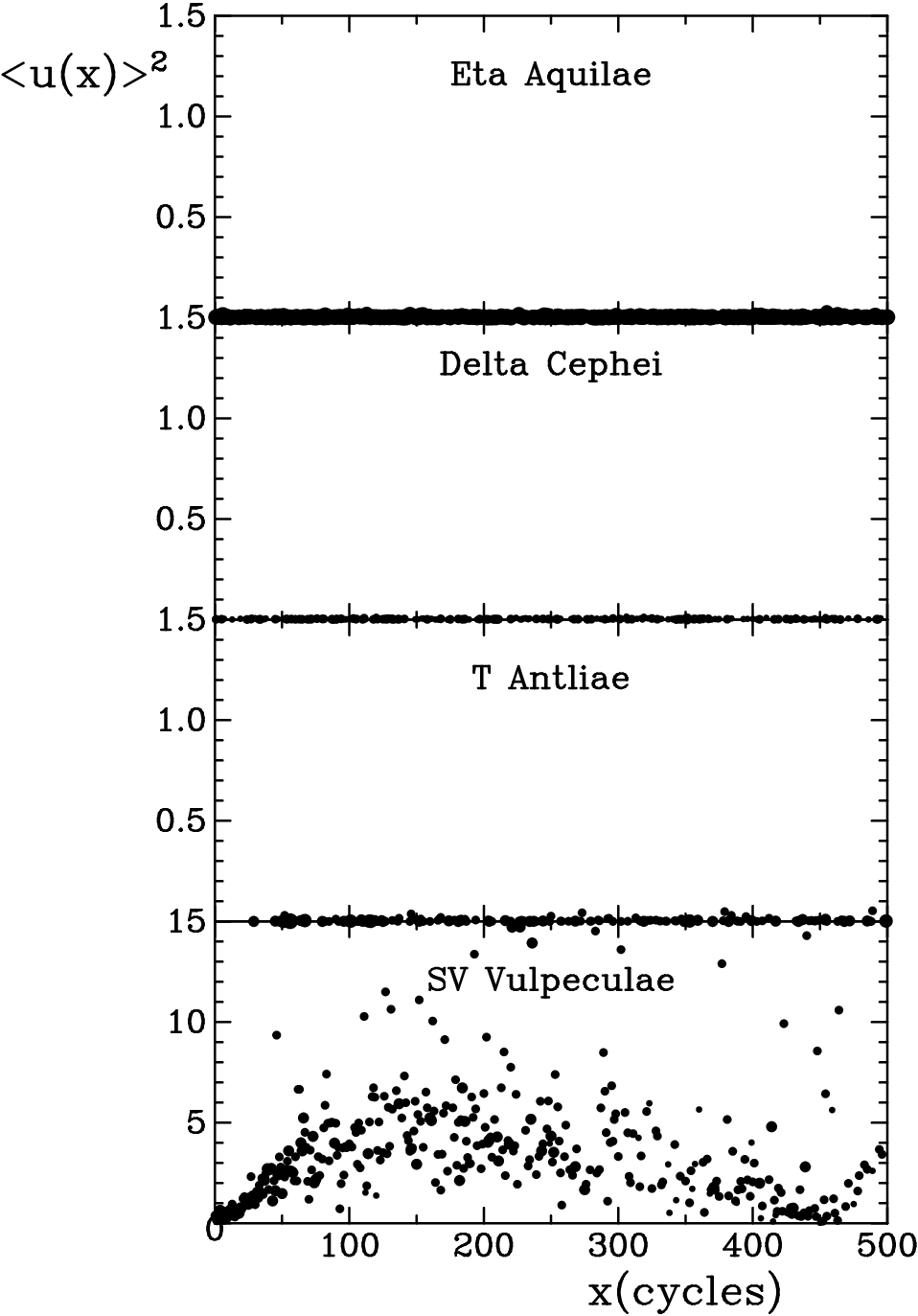}
\caption{\label{fig2} Eddington-Plakidis tests for four Cepheids.}
\end{figure}

Similar searches have been made using O--C data for Cepheid variables \citep{7,8,9,10,11,12,13}, but generally with negative results in most cases (see results for $\eta$ Aql, $\delta$ Cep, and T Ant in Fig.~\ref{fig2}). The long-period Cepheid SV Vul (Fig.~\ref{fig2}) represents the case of a positive detection of random fluctuations, and also a case where one can observe individual light maxima. The situation is complicated for short-period Cepheids because separate light maxima are rarely observed and individual times of maximum often refer to data obtained over many adjacent cycles about the one cited. Since random fluctuations in period exist over several pulsation cycles, their effects on times of light maximum can easily be confused with other sources of scatter in the light curves.

Similar evidence for short time scale random fluctuations in period can be seen in the results of \citet{14} for several LMC Cepheids. Offsets in the times of light maximum from cycle to cycle are clearly present in Poleski's O--C diagrams, but the deviations are relatively small in comparison with those observed over longer time intervals, where the effects of evolutionary changes in mean radius become dominant \citep{15}.

\begin{figure}[h]
\includegraphics[width=0.45\textwidth]{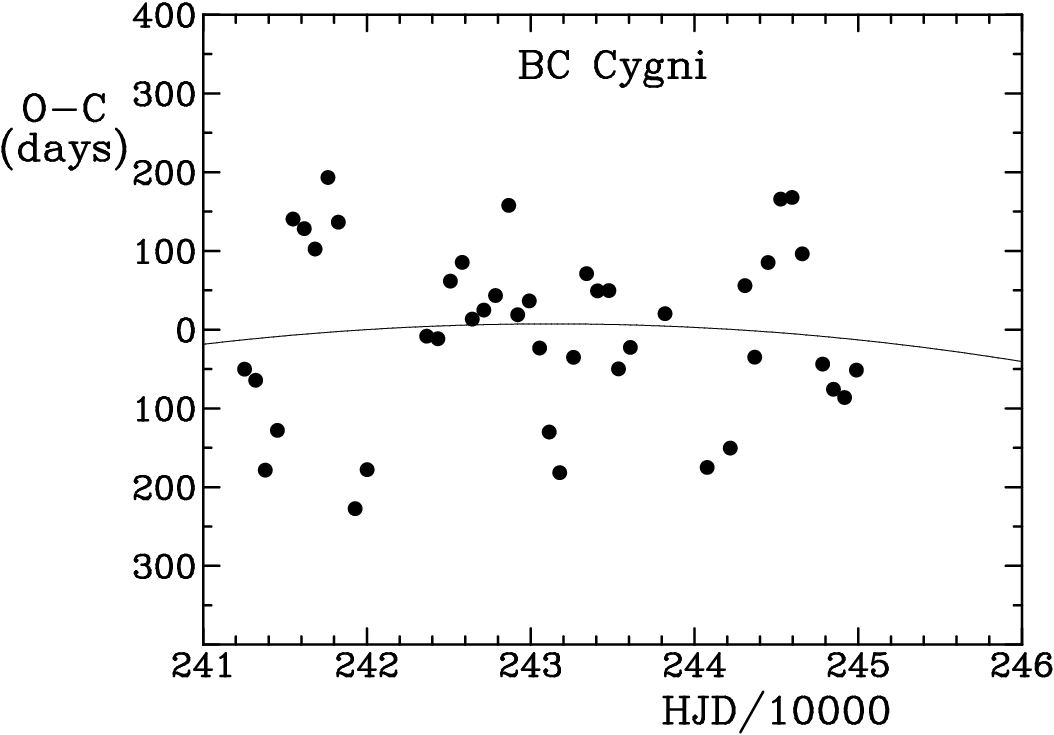}
\caption{\label{fig3} O--C trends for the M supergiant BC Cyg.}
\end{figure}

The M supergiant variable BC Cyg (M3 Ia) is an interesting example of random fluctuations in period for the SRC class of long-period variables, a small group of pulsating stars often overlooked by variable star observers. Times of light maxima for BC Cyg can generally be established to within a week or so ($\sim0.01 P$), but the O--C variations (Fig.~\ref{fig3}) display scatter amounting to as much as a few hundred days. The best-fitting downwards sloped parabolic trend is otherwise consistent with the established period decrease in BC Cyg between 1900 and 2000 \citep{16}.

\begin{figure}[h]
\includegraphics[width=0.45\textwidth]{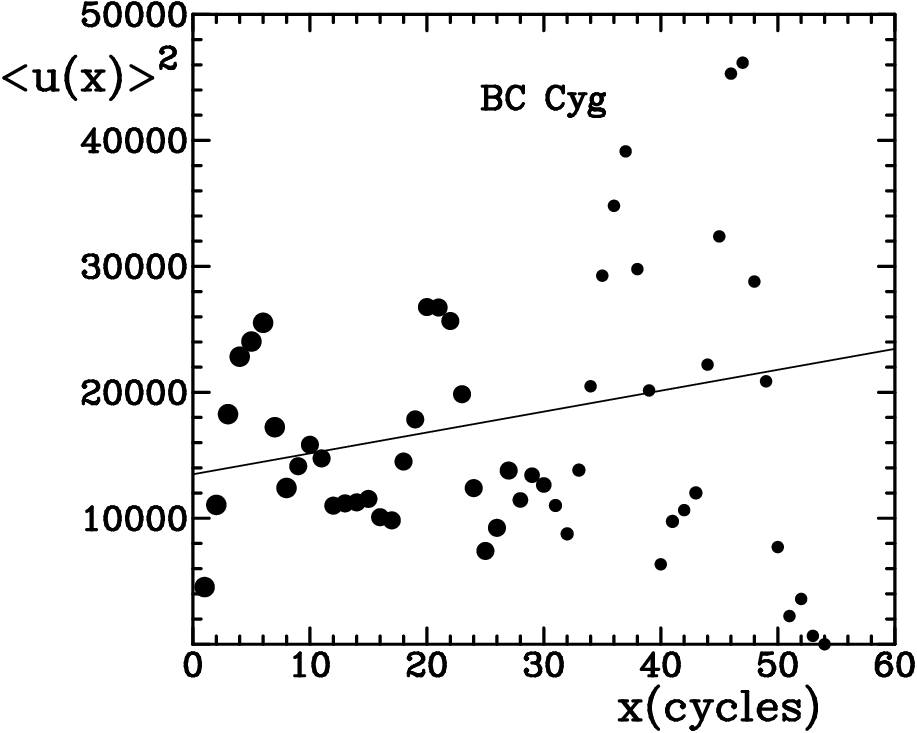}
\caption{\label{fig4} An Eddington-Plakidis test for BC Cyg.}
\end{figure}

\begin{figure}[h]
  \includegraphics[width=0.45\textwidth]{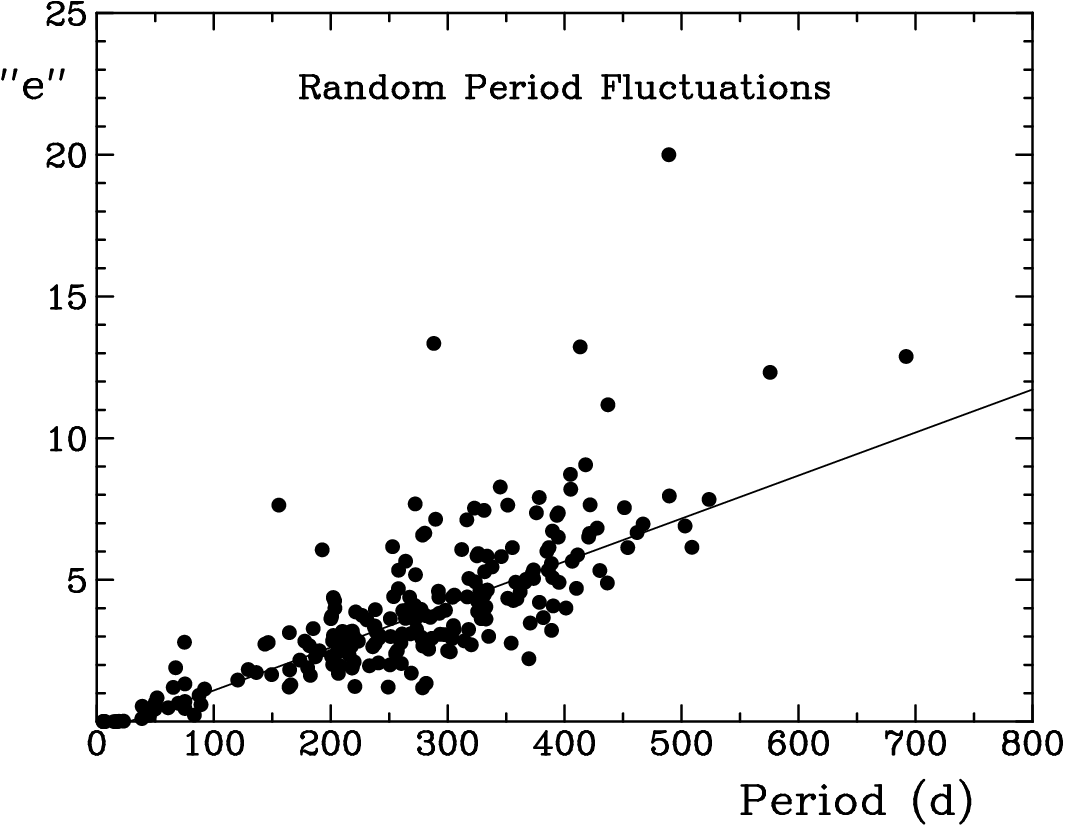}
  \caption{\label{fig5} The trend of increasing stochasticity {\it e} with pulsation period.}
\end{figure}

A test of the O--C deviations for BC Cyg using the Eddington-Plakidis technique is displayed in Fig.~\ref{fig4}. The deduced randomness parameter in this case of $e = 12.88$ is consistent with the recognized dependence of the parameter on pulsation period {\it P}. In general, all existing results suggest that the ``e'' parameter increases with period {\it P} \citep{7}, as shown in Fig.~\ref{fig5}. A working relationship for the dependence is given by:
\begin{displaymath}
e = -0.421 (\pm 0.146) + 0.015 (\pm0.001) P \;.
\end{displaymath}

\section{The e/P Parameter for Stochastic Changes}
\begin{figure}[h]
\includegraphics[width=0.45\textwidth]{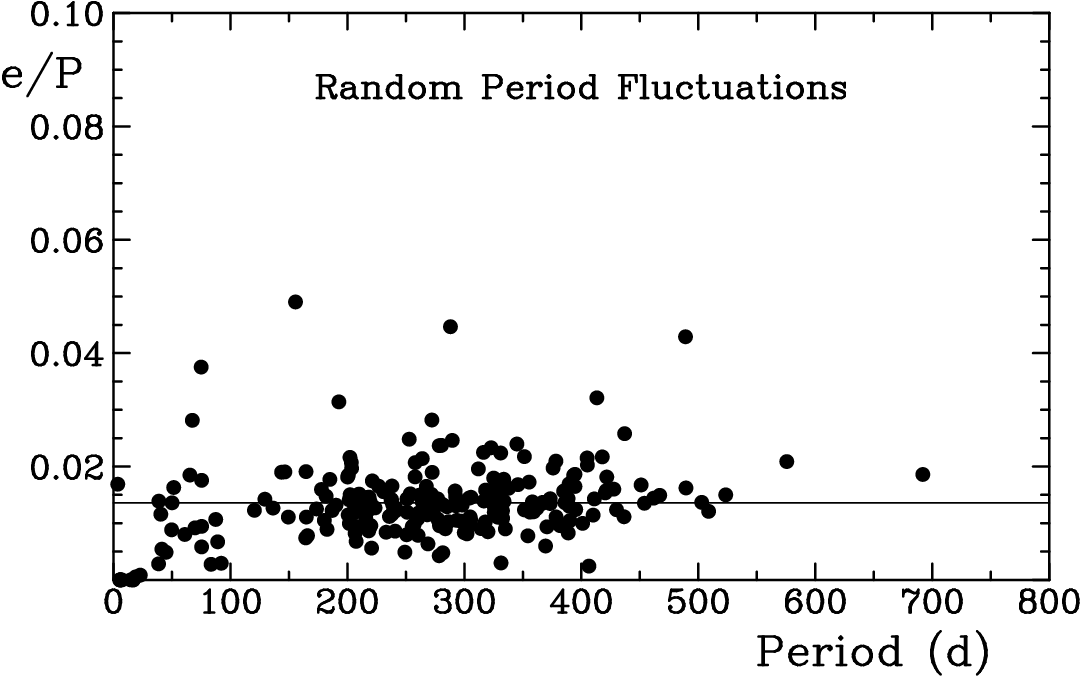}
\caption{\label{fig6} The ratio $e/P$ appears to be independent of pulsation period.}
\end{figure}

It was noted by \citet{1} that the observed random fluctuations in period for $o$ Cet (Mira) and $\chi$ Cyg amounted to 1.35--1.39\% of the pulsation period, a parameter that in turn depends directly on stellar radius. In other words, a better parameter for describing stochastic processes in pulsating stars should be the ratio $e/P$, which must be independent of radius if the variables obey a period-radius relation. Results to date for all Eddington-Plakidis analyses of pulsating stars (shown in Fig.~\ref{fig6}) confirm that assumption. The parameter $e/P$ is indeed relatively independent of pulsation period, {\it i.e.}, independent of stellar radius, and has a mean value of $0.0136 \pm0.0005$ ($\pm0.0069$ s.d.), identical to what \citet{1} concluded 80 years ago.

\section{Conclusions}
In stars such as SV Vul (Fig.~\ref{fig2}) and BC Cyg (Fig.~\ref{fig4}) the computed values of $\langle u(x) \rangle^2$ initially increase directly in proportion to increasing differences in cycle count {\it x}, as predicted for a stochastic process \citep{1}. At larger cycle differences, however, the trend is reversed as the dominant pulsation in such stars reimposes its regularity in the observed times of light maximum. The observed values of $\langle u(x) \rangle$ therefore become very small for large cycle differences. Similar characteristics are observed in many different pulsating stars, and was also noted by \citet{1}. Stochastic fluctuations in period appear to be a common feature of nearly all pulsating stars, but they are invariably dominated by the regular pulsation in such stars. The physical processes responsible for such fluctuations are uncertain, but presumably they originate in temporal modifications of envelope convection in such stars.


\begin{thebibliography}{16}

\bibitem[Abdel-Sabour~Abdel-Latif(2004)]{13} Abdel-Sabour Abdel-Latif, M.\ 2004 \emph{Ph.D. Thesis}, Cairo University

\bibitem[Berdnikov et al.(2004)]{9} Berdnikov, L.~N., 
Samus, N.~N., Antipin, S.~V., Ezhkova, O.~V., Pastukhova, E.~N., 
\& Turner, D.~G.\ 2004, \pasp, 116, 536 

\bibitem[Berdnikov et al.(2007)]{10} Berdnikov, L.~N., 
Pastukhova, E.~N., Gorynya, N.~A., Zharova, A.~V., 
\& Turner, D.~G.\ 2007, \pasp, 119, 82 

\bibitem[Berdnikov et al.(2009a)]{11} Berdnikov, L.~N., 
Pastukhova, E.~N., Turner, D.~G., 
\& Majaess, D.~J.\ 2009, Astronomy Letters, 35, 175 

\bibitem[Berdnikov et al.(2009b)]{12} Berdnikov, L.~N., 
Henden, A.~A., Turner, D.~G., 
\& Pastukhova, E.~N.\ 2009, Astronomy Letters, 35, 406 

\bibitem[Eddington 
\& Plakidis(1929)]{1} Eddington, A.~S., \& Plakidis, S.\ 1929, \mnras, 90, 65 

\bibitem[Percy et al.(1997)]{2} Percy, J.~R., Bezuhly, 
M., Milanowski, M., \& Zsoldos, E.\ 1997, \pasp, 109, 264 

\bibitem[Percy 
\& Hale(1998)]{3} Percy, J.~R., \& Hale, J.\ 1998, \pasp, 110, 1428 

\bibitem[Percy 
\& Colivas(1999)]{4} Percy, J.~R., \& Colivas, T.\ 1999, \pasp, 111, 94 

\bibitem[Percy et al.(2003)]{5} Percy, J.~R., Velocci, 
V., \& Sterken, C.\ 2003, \pasp, 115, 626 

\bibitem[Percy et al.(2007)]{6} Percy, J.~R., Bandara, 
K., \& Cimino, P.\ 2007, Journal of the American Association of Variable Star Observers (JAAVSO), 35, 343 

\bibitem[Poleski(2008)]{14} Poleski, R.\ 2008, Acta 
Astronomica, 58, 313 

\bibitem[Turner 
\& Berdnikov(2001)]{7} Turner, D.~G., \& Berdnikov, L.~N.\ 2001, Odessa Astronomical Publications, 14, 170 

\bibitem[Turner 
\& Berdnikov(2004)]{8} Turner, D.~G., \& Berdnikov, L.~N.\ 2004, \aap, 423, 335 

\bibitem[Turner et al.(2006a)]{15} Turner, D.~G., 
Abdel-Sabour Abdel-Latif, M., \& Berdnikov, L.~N.\ 2006, \pasp, 118, 410 

\bibitem[Turner et al.(2006b)]{16} Turner, D.~G., 
Rohanizadegan, M., Berdnikov, L.~N., 
\& Pastukhova, E.~N.\ 2006, \pasp, 118, 1533 


\end{thebibliography}
\end{document}